\documentclass[fleqn,10pt,dvipsnames]{article}
\usepackage[utf8]{inputenc}
\usepackage[T1]{fontenc}

\usepackage[left=2cm,%
  right=2cm,%
  top=2.25cm,%
  bottom=2.25cm,%
  headheight=12pt,%
  a4paper]{geometry}%

\usepackage{titling}
\usepackage{authblk}
\usepackage{cite}
\usepackage{amssymb,mathtools}  
\usepackage{bm}
\usepackage{physics2}
\usephysicsmodule{ab, ab.legacy, op.legacy}
\usepackage{fixdif, derivative}
\usepackage{xcolor}
\usepackage{booktabs, tabularx, nicematrix, makecell}
\usepackage{siunitx}
\usepackage{hyperref}
\usepackage{comment}

\newcommand{\rmsub}[2]{#1_\mathrm{#2}}
\newcommand{\rmsup}[2]{#1^\mathrm{#2}}
\DeclareMathOperator{\sign}{sign}

\newcommand{\xB}{\rmsub{\vec{x}}{B}}
\newcommand{\xP}{\rmsub{\vec{x}}{P}}
\newcommand{\xR}{\rmsub{\vec{x}}{R}}
\newcommand{\xM}{\rmsub{\vec{x}}{M}}
\newcommand{\xD}{\rmsub{\vec{x}}{D}}

\newcommand{\panellabel}[1]{{\Large \sffamily \textbf{#1} \par}}

\newcommand{\experimentmethod}{Ball possession game experiments}
\newcommand{\modeldetail}{Details on the dynamical model}
\newcommand{\nummethod}{Numerical methods and parameters}
\newcommand{\datagen}{Generating simulated dataset}
\newcommand{\sensanal}{Sensitivity analysis}

\newcommand{\siregtable}{Supplementary Table S1}
\newcommand{\siregcomp}{Supplementary Table S2}
\newcommand{\siexpvideo}{Supplementary Video S5}
\newcommand{\sisimvideowguide}{Supplementary Video S1}
\newcommand{\sisimvideo}{Supplementary Videos S1, S2, S3, and S4}
\newcommand{\sihistorg}{Supplementary Figure S1}
\newcommand{\sihistpro}{Supplementary Figure S2}

\title{%
Uncovering influence of football players' behaviour on team performance in ball possession through dynamical modelling 
}

\author[1,*]{Hidemasa Ishii}
\author[2]{Yohei Takai}
\author[1]{Yuichiro Marui}
\author[3]{Yoshihiro Yamazaki}
\author[1]{Yusuke Kato}
\author[1]{Hiroshi Kori}

\affil[1]{Department of Complexity Science and Engineering, Graduate School of Frontier Sciences, the University of Tokyo, Chiba 277-8561, Japan}
\affil[2]{National Institute of Fitness and Sports in Kanoya, Kagoshima 891-2393, Japan}
\affil[3]{Faculty of Science and Engineering, School of Advanced Science and Engineering, Waseda University, Tokyo 169-8555, Japan}

\affil[*]{hidemasaishii1997@g.ecc.u-tokyo.ac.jp}

\date{}

\begin{document}
\maketitle

\begin{abstract}
  A quest for uncovering influence of behaviour on team performance involves understanding individual behaviour, interactions with others and environment, variations across groups, and effects of interventions.
  Although insights into each of these areas have accumulated in sports science literature on football, it remains unclear how one can enhance team performance.
  We analyse influence of football players' behaviour on team performance in three-versus-one ball possession game by constructing and analysing a dynamical model.
  We developed a model for the motion of the players and the ball, which mathematically represented our hypotheses on players' behaviour and interactions.
  The model's plausibility was examined by comparing simulated outcomes with our experimental result.
  Possible influences of interventions were analysed through sensitivity analysis, where causal effects of several aspects of behaviour such as pass speed and accuracy were found.
  Our research highlights the potential of dynamical modelling for uncovering influence of behaviour on team effectiveness.
\end{abstract}

\section*{Introduction}  

Groups of humans coordinate their behaviour to achieve shared objectives.
We encounter teams of people coordinating for common causes in every aspect of our lives: people work together to raise families, serve dishes at restaurants, treat patients at hospitals, and publish scholarly articles.
Accordingly, team effectiveness has drawn significant attention, particularly in psychology, with a primary focus on approaches to enhancing team performance~\cite{kozlowski2006Enhancing,mathieu2019Embracing}.
Team sports including football are prominent cases where team effectiveness carries significance.
In sports, team performance, e.g. win-loss or scores, would be an obvious outcome of team effectiveness at the group level~\cite{mcewan2014Teamwork}.
While studies on `team effectiveness' in the context of sports remain scarce~\cite{mcewan2014Teamwork}, large literature in sports science has studied evaluation and enhancement of team performance in football.
A full football match is highly complex, involving diverse situational variables --- such as match location and league --- and interactions among 22 players, referees, coaches, and possibly the audience.
Moreover, a number of diverse events take place during a match, including passes, free kicks, and fouls, each of which may require a distinct description.
We therefore concentrate on a specific aspect of football: ball possession.
In particular, our study concerns three-versus-one (3v1) ball possession game, where three offensive players (OFs) possess the ball and one defensive player (DF) tries to intercept it.
In this game, OFs try to maximise the number of consecutive passes, while DF tries to minimise it.
Therefore the number of passes is the measure for performance.
Ball possession game, also known as rondo drill, is widely used in practice, although scarcely investigated~\cite{menuchi2018Effects,silva2024Acceleration}.
Ball possession is one of fundamental elements in football, since a match consists of alternating periods of possessions by two opposing teams.
Among the literature on performance analysis, it has been one of the most studied indicators of team performance~\cite{collet2013possession,bradley2013effect}, and its relationship with situational variables~\cite{lago-penas2010Ball,merlin2020Exploring,maneiro2021Identification,bradley2014influence,bransen2024datadriven} and other performance indicators~\cite{bradley2013effect,garcia-calvo2023How,jerome2024Evidence} has been extensively analysed.
Recently, it also attracts attention in physics and computer science~\cite{chacoma2020Modeling,yamamoto2024Theory}.
The accumulated insights suggest that ball possession is strongly associated with aggregated team performance, while it is not necessarily predictive of individual match results~\cite{collet2013possession,wang2022systematic}.

Regarding football, research in sports performance analysis has accumulated insights into players' behaviour, aiming to enhance individual and team performance~\cite{mcgarry2009Applied,mackenzie2013Performance}.
The literature has repeatedly highlighted the significance of clarifying the relationship between players' behaviour and team performance~\cite{mcgarry2009Applied,travassos2013Performance,glazier2017Grand,wunderlich2024Using}.
A quest for uncovering influence of players' behaviour on performance involves four key areas of focus:
(1) individual players' behaviour, 
(2) their interactions with others and the environment, 
(3) variations in their behaviour across groups, and 
(4) the effects of interventions in behaviour on performance.
Each of the four areas has been studied in the literature on football.
Regarding individual behaviour, previous research considered diverse factors including sprint motion~\cite{fujimura2005Geometric,narizuka2023Validation}, the change of direction~\cite{deutsch2024Frequency}, and anthropometric, physical, psychological and technical characteristics~\cite{taylor2005Comparison,rebelo2012Anthropometric,koudellis2024Physical,bastiere2024Active}.
Since football is a team sport, interactions among players have attracted considerable attention.
The literature on intra-team interactions focused on formation~\cite{frencken2011Oscillations,laakso2019Interpersonal,kunrath2024Youth}, passing networks~\cite{duch2010Quantifying,narizuka2014Statistical,yamamoto2021Preferential,pan2024Evolution}, and motion alignment among teammates~\cite{yokoyama2011Three,folgado2014Competing,yokoyama2018Social,chacoma2021Stochastic}.
Interactions with the opposing team have also been analysed~\cite{narizuka2016Statistical,chacoma2022Complexity,modric2024It}.
Players' response to the environment, which is often referred to as situational variables~\cite{sarmento2018Influence}, has been a central topic as well.
Previous studies examined influences of temporal~\cite{sarmento2018Influence,pan2024Evolution,bransen2024datadriven}, spatial~\cite{sarmento2018Influence,menuchi2018Effects,laakso2019Interpersonal}, and strategic~\cite{torrents2016Emergence,sarmento2018Influence,demetrio2024effects,kunrath2024Youth} factors.
Variations in behaviour across groups have been actively studied, with groups being defined in various ways, including competitive levels~\cite{rebelo2012Anthropometric,bradley2013effect,collet2013possession,yokoyama2018Social,koudellis2024Physical,piechota2024Expert}, field positions~\cite{rebelo2012Anthropometric,laakso2019Interpersonal,castellano2024Match,deutsch2024Frequency,modric2024It}, and confederations~\cite{ju2024Analysis}.
Finally, in the literature on effects of interventions, one finds two major approaches.
The first is a direct approach through experiments, as seen in the research on training programs~\cite{arslan2024effects,sal-de-rellan2024Effects}.
The other approach involves comparison between competent and less competent players. 
These studies aim to extract characteristics that are associated with performance, which are sometimes called performance variables~\cite{travassos2013Performance}.
Associations between explanatory variables and performance are considered to be suggestive of causal effects~\cite{piechota2024Expert,koudellis2024Physical}.
Despite these contributions on each area, studies examining their interplay remain scarce.
Furthermore, it has been difficult to discuss causality between behaviour and performance.
A difference in a certain quantity does not imply its causal effect.
That is, even if a certain quantity differs between players at different competitive levels, this difference may well not be the origin of the difference in performance.
Indeed, Yokoyama et al. reported an interesting result from their experiment: although they succeeded in modifying novice players' motion to exhibit a similar synchronisation pattern to that of the experienced, the performance of the novice did not improve~\cite{yokoyama2018Social}.
It suggests the possibility of spurious correlation between motion synchronisation patterns and teams' competence.
Due to these limitations, it still remains unclear how one can enhance team performance in football.

In this research, we analyse influence of football players' behaviour on ball possession performance by constructing and analysing a dynamical model.
Building on a previous work by Yokoyama et al.~\cite{yokoyama2018Social}, we developed a dynamical model describing the motion of the players and the ball in 3v1 ball possession game.
There are three advantages in our modelling approach.
Firstly, the use of dynamical models is a direct and natural strategy to take dynamic and emergent aspects of team-level processes into consideration.
Among the literature on team effectiveness, it has been argued that, although conventional team effectiveness research has examined team processes in static frameworks, more focus should be put on their dynamic, complex, and emergent aspects~\cite{kozlowski2018Unpacking,mathieu2019Embracing}.
Team processes dynamically arise through microscopic interactions among team members.
Hence, there is a need for methodologies that can incorporate the dynamic interplay among individual behaviour and their interactions.
Secondly, the modelling approach spans the four key areas of focus mentioned above.
That is, the use of dynamical models allows us to study not only each of the four areas but also the interplay among them, as we demonstrate in this article.
A model mathematically represents hypotheses about individual behaviour (area 1) and interactions with others and with the environment (area 2).
Variations in behaviour across groups (area 3) can be described by different parameter values of the model.
Effects of interventions (area 4) can be analysed through sensitivity analysis, where effects of model parameters are examined.
Lastly, our approach has another significant advantage: it allows us to discuss causal effects of behaviour on performance by drawing a clear distinction between behaviour and outcome.
In this research, we consider that model parameters, each of which controls a certain aspect of players' motion, characterise behaviour.
A simulation of the model generates time series data of positions of the players and the ball, similarly to a tracking system.
Various outcome quantities, including performance, can be calculated from the simulated position data.
This means that we can control a specific aspect of players' behaviour in our model, measure its effect on outcomes, and discuss causality between behaviour and outcome.
We note that our approach shares motivations with what is referred to as dynamical systems approach~\cite{frencken2011Oscillations,duarte2012Sports,glazier2017Grand,rein2017Maybe} or ecological dynamics approach~\cite{travassos2013Performance,menuchi2018Effects,laakso2019Interpersonal} in sports science.
Studies adopting these approaches typically conduct statistical analyses with empirical data in terms of quantities that reflect the temporal and interacting nature of team dynamics, such as distance between players at each time frame and cross-correlation among players.
Nevertheless, our research strategy is distinct from these approaches in that we construct a dynamical model for our specific purpose, as has been done in the physics and computer science literature related to football~\cite{narizuka2014Statistical,narizuka2016Statistical,yokoyama2018Social,chacoma2020Modeling,chacoma2021Stochastic,chacoma2022Complexity,narizuka2023Validation}.

In the remainder of this article, we first report the experimental result of 3v1 ball possession games.
The participants were grouped into high- and low-level teams.
We compared the two teams based on the motion of OFs, quantified by the offensive team area (OF area), and their performance, measured by the number of successful passes.
We then present an overview on the constructed model which describes the motion of the players and the ball.
The model represents our hypotheses about behaviour of players and their interactions.
In the next subsection, we evaluate the plausibility of our model by comparing the two outcome quantities, OF area and the number of passes, with the experiment.
Finally, possible influence of interventions in behaviour on ball possession performance is examined through sensitivity analysis.
Our work provides researchers studying team sports, in particular team performance in football, with a model for 3v1 ball possession game whose details can be adjusted according to their interest.
Moreover, we illustrate the potential of the modelling approach for investigating causal effects of certain aspects of behaviour on team effectiveness in broader contexts.

\section*{Results}
\begin{figure}[tp]
  \centering
  \includegraphics[width=0.96\linewidth]{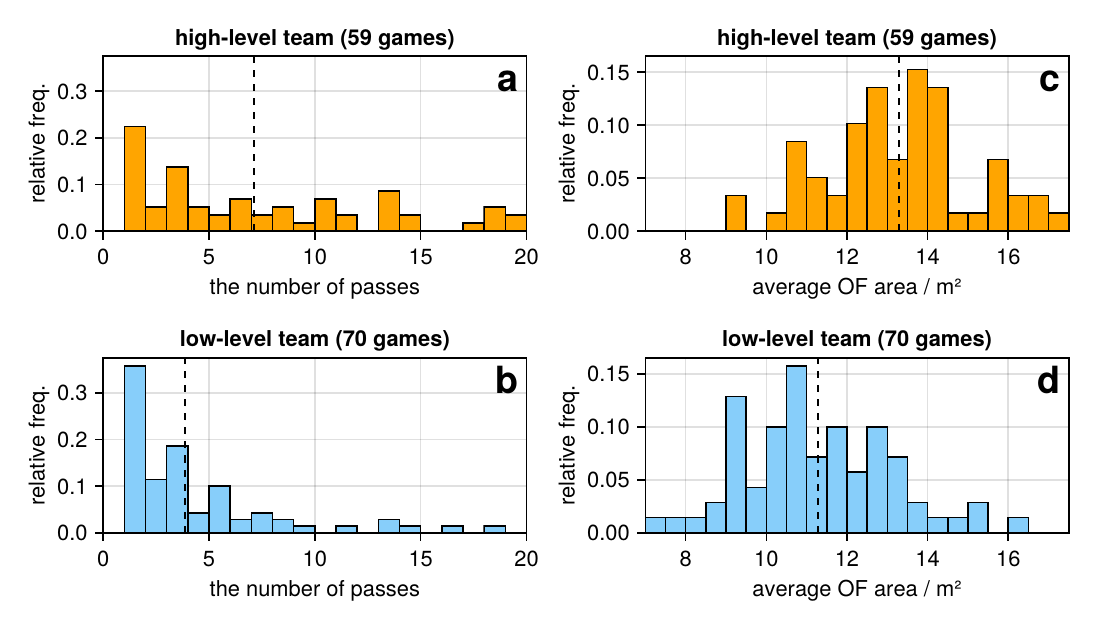}
  \caption{
    Comparison of experimental results between high- and low-level teams.
    The histograms show empirical distributions of 
    (a) the number of passes for the high-level team,
    (b) the number of passes for the low-level team,
    (c) the average OF area, i.e. the average area of a triangle formed by OFs, for the high-level team, and 
    (d) the average OF area for the low-level team.
    A vertical dashed line in each panel indicates the mean value.
  }
  \label{fig:emp.dist}
\end{figure}

\begin{figure}[tp]
  \newcommand{\subitemsymb}{---}
  \setlength{\tabcolsep}{3pt}
  \newlength{\btwspacing}
  \setlength{\btwspacing}{4pt}
  \centering
  \begin{tabular}{c}
    \begin{minipage}{.96\linewidth}
      \panellabel{a}
      \centering
      \includegraphics[width=0.96\linewidth]{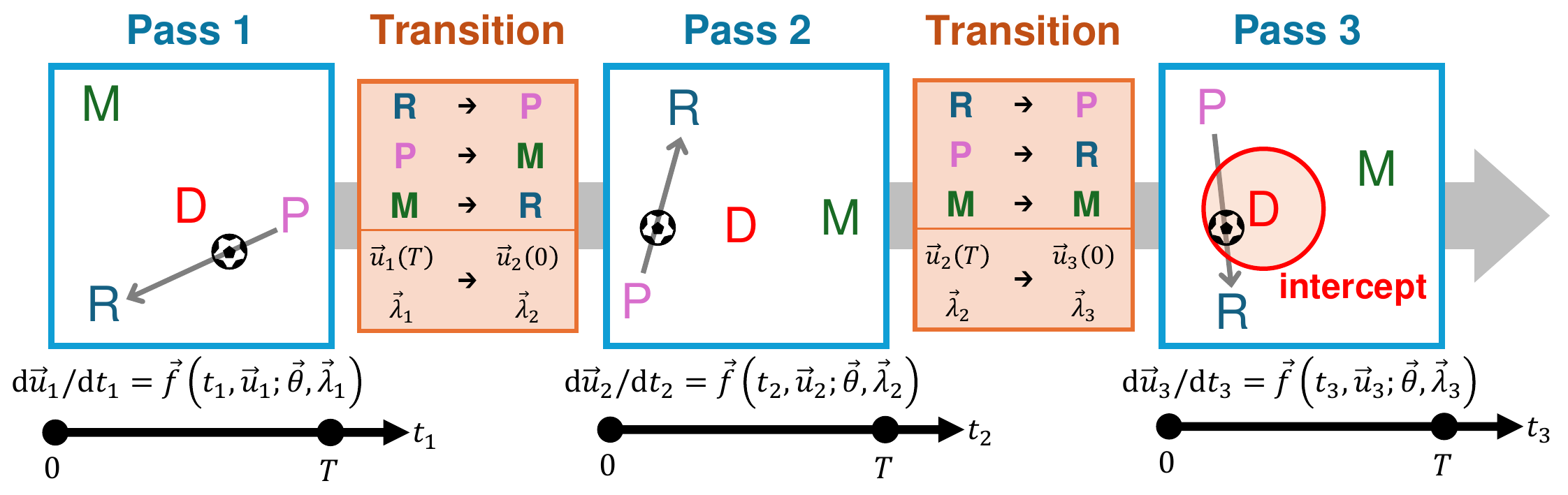}
    \end{minipage}
    \\[3cm]
    \begin{minipage}{.96\linewidth}
      \small
      \panellabel{b} \vspace{6pt}
      \centering
      \begin{tabular}{rrl}
        \toprule
        1. & \multicolumn{2}{l}{Initialise the system} \\
        & \subitemsymb & Set the pass count to $1$ \\
        & \subitemsymb & Determine the initial condition of the state vector $\vec{u}_1(t_1)$ and the pass-specific parameters $\vec{\lambda}_1$ \\[\btwspacing]
        2. & \multicolumn{2}{l}{Simulate dynamics during a single pass scene} \\
        & \subitemsymb & During the $n$-th pass scene, $\vec{u}_n(t_n)$ evolves according to a system of ODEs~\eqref{eq:ode.system} with parameters $\vec{\theta}$ and $\vec{\lambda}_n$: \\
        && \hfill $\d \vec{u}_n / \d t_n = \vec{f}(t_n, \vec{u}_n; \vec{\theta}, \vec{\lambda}_n)$ for $t_n \in [0, T]$ \hfill\null \\[\btwspacing]
        3. & \multicolumn{2}{l}{Terminate the simulation if a termination condition has been satisfied} \\
        & \subitemsymb & Check three conditions: pass intercept, ball-out, and the maximum number of passes \\[\btwspacing]
        4. & \multicolumn{2}{l}{Transition to the next pass scene} \\
        & \subitemsymb & \makecell[lt]{
          Determine the next pass-specific parameters $\vec{\lambda}_{n+1}$ and the initial condition for $\vec{u}_{n+1}(t_{n+1})$ stochastically \\ based on the current parameters $\vec{\lambda}_{n}$ and the final state of the current pass scene $\vec{u}_{n}(T)$
        } \\
        & \subitemsymb & Update the pass count from $n$ to $n+1$ \\[\btwspacing]
        5. & \multicolumn{2}{l}{Go back to step 2} \\
        \bottomrule
      \end{tabular}
    \end{minipage}
  \end{tabular}
  \caption{
    Overview of our model.
    (a) A game comprises a sequence of pass scenes and transitions between successive pass scenes.
    (b) Our model breaks down into five steps, through which the state vector during the $n$-th pass scene, $\vec{u}_n(t_n)$, and the pass count evolve.
  }
  \label{fig:model.overview}
\end{figure}

\begin{table}[tp]
  \centering
  \caption{
    List of model parameters and their value ranges in this study.
    The entry of --- in `unit' column stands for a dimension-less quantity.
  }
  \label{tab:param.list}
  \begin{tabular}{rrrl}
    \toprule
    parameter & value range & unit & description \\
    \midrule
    $m$ & 55 -- 75 & $\unit{kg}$ & Mass of each player. \\
    $T$ & 0.5 -- 1.2 & $\unit{s}$ & Duration of every pass scene. \\
    $\sigma$ & 3 -- 9 & $\unit{\radian}$ & Standard deviation of pass angles. \\
    $\beta$ & 0 -- 25 & --- & Accuracy for Passer to choose the wider side in giving a pass. \\
    $\rmsub{k}{r}$ & 0 -- 1000 & $\unit{kg.m/s^2}$ & Spring constant of the returning force. \\
    $\gamma$ & 0 -- 1000 & $\unit{kg.s/m^2}$ & Viscosity of the neighbourhood of the court boundary. \\
    $\rmsub{k}{f}$ & 0 -- 200 & $\unit{kg/s^2}$ & Spring constant of the following force. \\
    $\rmsub{L}{f}$ & 3 -- 8 & $\unit{m}$ & Natural length of the following force. \\
    $\rmsub{k}{e}$ & 0 -- 200 & $\unit{kg/s^2}$ & Spring constant of the evading force. \\
    $\rmsub{L}{e}$ & 1 -- 6 & $\unit{m}$ & Natural length of the evading force. \\
    $q$ & 0.2 -- 0.8 & --- & \makecell[lt]{Probability that DF chooses the same motion mode as in the previous \\ pass scene.} \\
    $\tau$ & 0.5 -- 1.5 & $\unit{s}$ & Time scale of DF's motion. \\
    $\rmsub{L}{ict}$ & 0.4 -- 0.8 & $\unit{m}$ & Threshold distance for DF to intercept the ball. \\
    $\rmsub{L}{buff}$ & $0.5$ & $\unit{m}$ & Width of the viscous region around the court boundary. \\
    $\rmsub{L}{out}$ & $0.5$ & $\unit{m}$ & Threshold distance for evaluating the `ball-out' termination condition. \\
    $L$ & $6$ & $\unit{m}$ & Length of each side of the square court.
    \\ \bottomrule
  \end{tabular}
\end{table}

\begin{figure}[tp]
  \centering
  \includegraphics[width=0.96\linewidth]{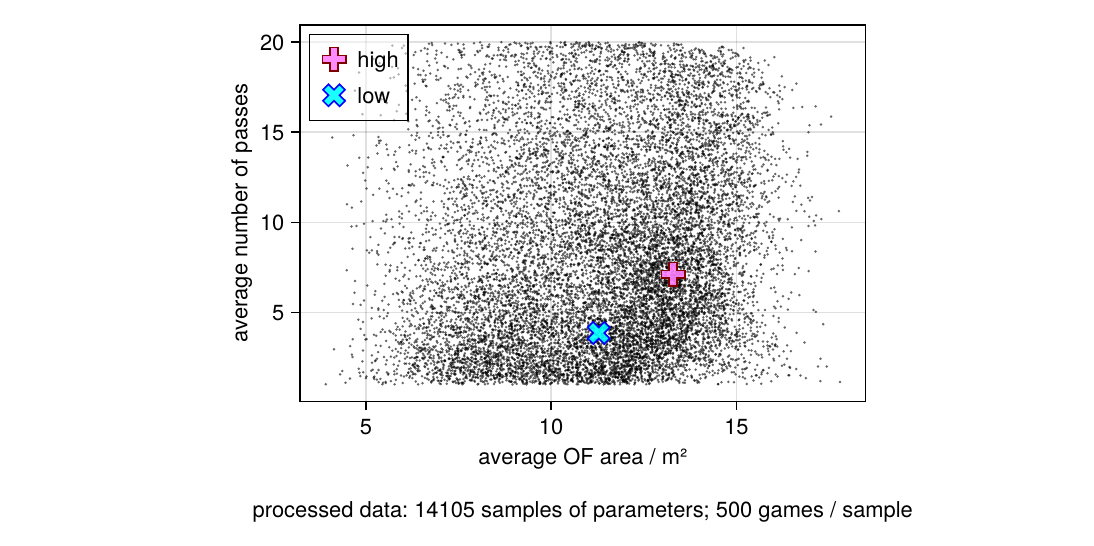}
  \caption{%
    Comparison between empirical and simulated datasets.
    Each black point represents the ensemble average of the average OF area and the number of passes over 500 simulated games for a set of randomly chosen parameter values.
    The simulated dataset covered the empirical data for both high- and low-level teams (cross markers).
    The simulated dataset contained 14,105 samples that remained after the preprocessing.
  }
  \label{fig:random.params}
\end{figure}

\begin{figure}[tp]
  \centering
  \includegraphics[width=0.96\linewidth]{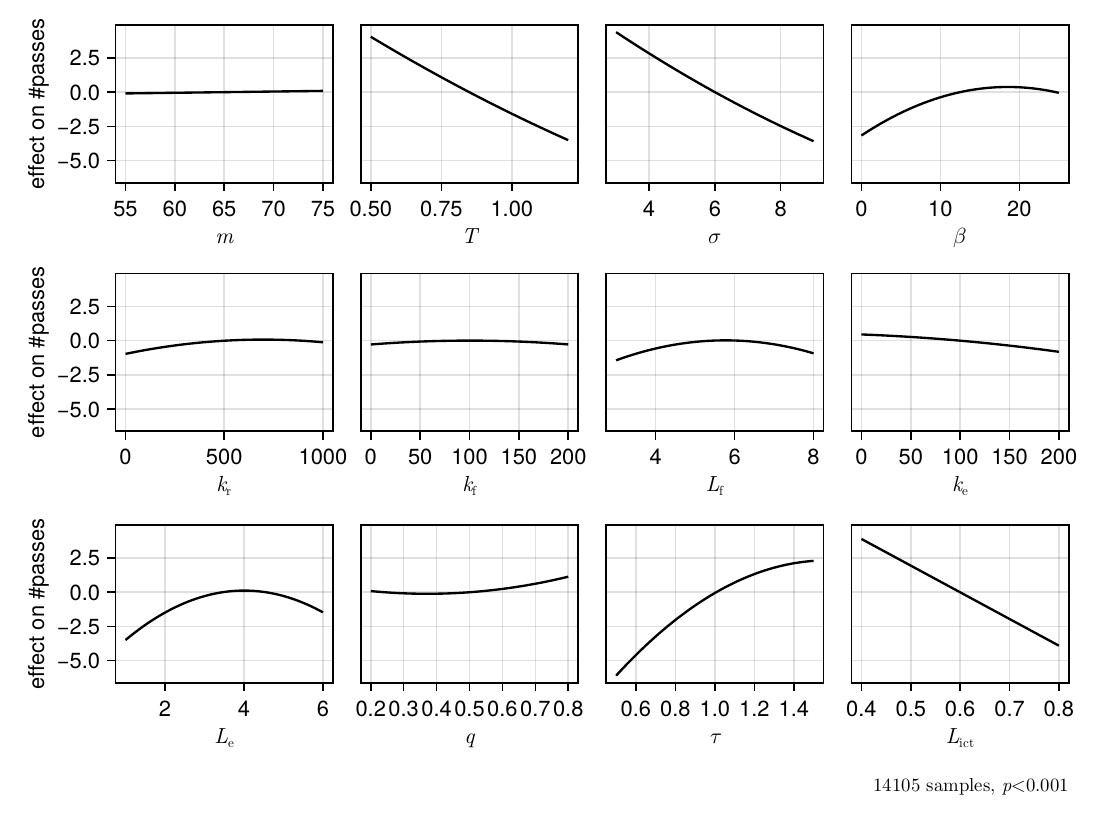}
  \caption{%
    Statistically significant ($0.1\%$) quadratic effect of each model parameter on the number of passes.
    The value of $0$ on $y$-axis corresponds to the reference value which is the prediction of the fitted regression model~\eqref{eq:quadratic.model} when all the parameters assume their sample average values within the simulated dataset.
    Each panel shows the predicted deviation in the number of passes from the reference value against the change in a parameter value ($x$-axis), when the other parameters are kept at their means.
    We refer readers to `\sensanal{}' in Methods section for methodological details.
  }
  \label{fig:sens.anal}
\end{figure}

\begin{table}[tp]
  \centering
  \caption{%
    Statistically significant effects of model parameters with the significance level of $0.1\%$.
    For linear and square effects, directions of influences, i.e. signs of coefficients, are indicated.
    For interaction effects, each entry shows the number of interaction terms whose coefficients were significant.
    Since we put 13 parameters as independent variables, there were 12 interaction terms for each parameter.
  }
  \label{tab:param.effect}
  \renewcommand{\arraystretch}{1.2}
  \setlength{\tabcolsep}{4pt}
  \begin{tabular}{rrrrrrrrrrrrrrr}
    \toprule
    type of effect & $m$ & $T$ & $\sigma$ & $\beta$ & $\rmsub{k}{r}$ & $\gamma$ & $\rmsub{k}{f}$ & $\rmsub{L}{f}$ & $\rmsub{k}{e}$ & $\rmsub{L}{e}$ & $q$ & $\tau$ & $\rmsub{L}{ict}$ \\
    \midrule
    linear ($b_k$) & $+$ & $-$ & $-$ & $+$ & $+$ & n.s. & n.s. & $+$ & $-$ & $+$ & $+$ & $+$ & $-$ \\
    square ($c_k$) & n.s. & $+$ & $+$ & $-$ & $-$ & n.s. & $-$ & $-$ & $-$ & $-$ & $+$ & $-$ & n.s. \\
    interaction ($d_{kl}$) & 1/12 & 7/12 & 8/12 & 6/12 & 6/12 & 0/12 & 2/12 & 8/12 & 6/12 & 9/12 & 5/12 & 8/12 & 7/12
    \\ \bottomrule
  \end{tabular}
\end{table}

\subsection*{Empirical differences between high- and low-level teams}
Fig.~\ref{fig:emp.dist} shows the result of the 3v1 ball possession game experiments.
Participants, all of whom had over 10 years of competitive experience, were grouped into high- and low-level teams.
They performed $59$ ($70$, respectively) ball possession games among the high-level (low-level) team members.
To characterise the motion of the offensive team, the temporal average of OF area, which was the area of the triangle formed by OFs, was recorded for each game.
The number of consecutive passes was recorded as well to measure team performance.
Comparison between Fig.~\ref{fig:emp.dist}a (high-level) and Fig.~\ref{fig:emp.dist}b (low-level) indicates the higher competence of the high-level team.
That is, the team performed better than the low-level one on average.
Panels c and d (Fig.~\ref{fig:emp.dist}) indicate that the two teams also differed with regard to their motion represented by the OF area.
The high-level team maintained a larger formation on average.
The extended formation is intuitively expected to lower the probability of intercepts, because then DF has to run more intensely to reach the ball.
Motivated by these observations, we proceeded to construct a model capable of reproducing the characteristics of both teams.

\subsection*{Dynamical model for three-versus-one ball possession game}
We constructed a dynamical model for 3v1 ball possession game, which is summarised in Fig.~\ref{fig:model.overview}.
A single game consists of a sequence of pass scenes and transitions between successive passes (Fig.~\ref{fig:model.overview}a). The model governs the motion of five agents: the ball (B), one DF (D), and three OFs.
Among OFs, one is Passer (P) making a pass, another is Receiver (R) receiving the ball, and the other is called Mover (M), who does not touch the ball during the pass scene.
The state variables of the system, which are positions and velocities of the agents, evolve according to the algorithm summarised in Fig.~\ref{fig:model.overview}b.

During each pass scene (step 2), motion of the players and the ball is described by the system of ordinary differential equations (ODEs).
That is, the state vector during the $n$-th pass scene $\vec{u}_n(t_n)$, which consists of positions $\vec{x}_j$ and velocities $\vec{v}_j$ of agent $j$, evolves according to the following ODE system:
\begin{subequations} \label{eq:ode.system}
\begin{align}
  \label{eq:B.x.dot}
  T \odv{\rmsub{\vec{x}}{B}}{t_n} =& \rmsub{\vec{d}}{B}^n - \rmsub{\vec{s}}{B}^n, \\
  \label{eq:R.x.dot}
  T \odv{\rmsub{\vec{x}}{R}}{t_n} =& \rmsub{\vec{d}}{B}^n - \rmsub{\vec{s}}{R}^n, \\
  \label{eq:DF.x.dot}
  \tau \odv{\rmsub{\vec{x}}{D}}{t_n} =& \bab{\zeta_n \rmsub{\vec{d}}{D, \text{chase}}(t_n) + \pab{1 - \zeta_n} \rmsub{\vec{d}}{D, \text{wait}}(t_n)}- \rmsub{\vec{x}}{D}, \\
  \intertext{and for $i \in \Bab{\mathrm{P}, \mathrm{M}}$,}
  \label{eq:PM.x.dot}
  \odv{\vec{x}_i}{t_n} =& \vec{v}_i, \\
  \label{eq:PM.v.dot}
  m \odv{\vec{v}_i}{t_n} =& \rmsup{\vec{F}}{return}(\vec{x}_i) + \rmsup{\vec{F}}{viscous}(\vec{x}_i, \vec{v}_i) + \vec{G}(t_n, \vec{u}_n; \vec{\theta}, \vec{\lambda}_n),
\end{align} \end{subequations}
where $t_n \in \bab{0, T}$ denotes time within the $n$-th pass scene, $\rmsub{\vec{s}}{B}^n$ and $\rmsub{\vec{s}}{R}^n$ are the starting position of the ball and Receiver respectively, $\rmsub{\vec{d}}{B}^n$ is the destination of the ball, $\zeta_n \in \Bab{0, 1}$ is an indicator of DF's motion mode (either `chase' or `wait'), and $\rmsub{\vec{d}}{D,mode}(t_n)$ determines the velocity of DF at time $t_n$ according to its motion mode.
Model parameters including $T$, $\tau$ and $m$ are described in table~\ref{tab:param.list}.
The dynamics of Passer and Mover [eqs.~\eqref{eq:PM.x.dot} and \eqref{eq:PM.v.dot}] involve social forces $\rmsup{\vec{F}}{return}(\vec{x}_i)$, $\rmsup{\vec{F}}{viscous}(\vec{x}_i, \vec{v}_i)$ and $\vec{G}(t_n, \vec{u}_n; \vec{\theta}, \vec{\lambda}_n)$.
The returning force $\rmsup{\vec{F}}{return}$ and the viscous force $\rmsup{\vec{F}}{viscous}$ constitute the confinement mechanism to keep players inside the court.
The remaining part $\vec{G}$ expresses hypotheses about players' behaviour and interactions.
For conciseness, we also denote the ODE system~\eqref{eq:ode.system} as
$\d \vec{u}_n / \d t_n = \vec{f}(t_n, u_n; \vec{\theta}, \vec{\lambda}_n)$.
We highlight that $\vec{f}$ takes two types of parameters:
constant model parameters $\vec{\theta}$ and pass-specific parameters $\vec{\lambda}_n$.
The vector of model parameters $\vec{\theta}$, which includes $T$, $\tau$ and $m$, is constant throughout a game.
In contrast, parameters in $\vec{\lambda}_n$, such as $\rmsub{\vec{s}}{B}^n$, $\rmsub{\vec{d}}{B}^n$ and $\zeta_n$, may assume different values across pass scenes.

Numerical integration of the ODE system~[eq.~\eqref{eq:ode.system}] in step 2 generates a time series of $\vec{u}_n(t_n)$ for $t_n \in \bab{0, T}$.
In step 3, termination conditions are checked, and the simulation is stopped if one of the conditions has been satisfied during the last pass scene.
A game ends in one of three situations: DF intercepts the ball, the ball goes out of the court, or the maximum number of passes, which is 20, is reached.
When no condition is fulfilled, the transition to the next pass scene takes place (step 4).
Pass-specific parameters of the $(n+1)$-th pass scene, such as the destination of the ball $\rmsub{\vec{d}}{B}^{n+1}$ and the DF's motion mode $\zeta_{n+1}$, are determined stochastically based on the state $\vec{u}_n$ and parameters $\vec{\lambda}_n$ of the $n$-th, i.e. preceding, pass scene.
The initial condition for the $(n+1)$-th pass scene also stochastically depends on the preceding pass scene.
These steps from 2 to 4 are repeated until one of the termination conditions is satisfied.
We refer readers to `\modeldetail{}' in Methods section for details on the model.

\subsection*{Plausibility of the model}
To examine the plausibility of the model, we conducted a large number of simulations and compared their results with the empirical data.
We sampled $15,000$ sets of parameter values randomly, and simulated $500$ games for each set of parameter values.
As in the experiments, for each game, the temporal average of OF area was calculated to characterise the motion of the offensive team, and the number of successful passes was also recorded to measure the team performance.
Their ensemble averages over $500$ trials are depicted as black points in Fig.~\ref{fig:random.params}.
We refer readers to `\datagen{}' in Methods section for details on data generation.
The red and blue cross markers in the figure denote the empirical values of the high- and low-level teams.
The simulation results covered the empirical values in terms of both the number of passes and the average OF area.
Therefore, we argue that the simulation results suggest the plausibility of our model in terms of both the motion and the performance of a team.
It is worth mentioning that not only the quantities measured here but also the dynamics during a simulated game (\sisimvideo{}) visually resembled empirical tracking data (\siexpvideo{}).

\subsection*{Effects of model parameters on the number of passes}
Taking advantage of the modelling approach, we examined determinants of strong performance, namely the larger number of passes, in our model.
To analyse the influence of each model parameter on the number of passes, we conducted sensitivity analysis using the dataset visualised in Fig.~\ref{fig:random.params}.
To be precise, we performed a multivariate quadratic regression whose dependent variable was the average number of passes and whose independent variables were 13 model parameters that had been varied.
We chose the quadratic regression, because we expected nonlinear effects of several parameters.
We refer readers to `\sensanal{}' in Methods section for details.
The result from the regression analysis is summarized in table~\ref{tab:param.effect}.
Moreover, the quadratic effect of each parameter, assuming all the other parameters assume their mean values, is visualized in Fig.~\ref{fig:sens.anal}.

While constructing the model, we expected several parameters to have limited or trivial influences on the number of passes.
One of such parameters was $m$, the mass of each player.
Since $m$ only affects the inertia of Passer and Mover [eq.~\eqref{eq:PM.v.dot}], we expect it to have a little effect on the number of passes.
Indeed, table~\ref{tab:param.effect} indicates that $m$ had little significant effect.
Although its linear coefficient turned out to be significant, Fig.~\ref{fig:sens.anal} suggests that its magnitude was negligible.
Parameters $\gamma$ and $\rmsub{k}{r}$ control the confinement mechanism.
Since the mechanism should have little influence over the dynamics inside the court, it is desirable that $\gamma$ and $\rmsub{k}{r}$ have little effect on the outcomes.
The sensitivity analysis revealed that $\gamma$ had no significant effect as expected.
As for $\rmsub{k}{r}$, although it exhibited a certain influence, Fig.~\ref{fig:sens.anal} suggests that its magnitude was limited.
Parameters $T$ and $\sigma$ are expected to reflect the competence of the offensive team.
Parameter $T$ is the duration of every pass scene fixed throughout a game.
A smaller value of $T$, which implies faster passes, should reflect the higher competence of the team.
Parameter $\sigma$ is the standard deviation of the pass angle.
A smaller value of $\sigma$ results in more accurate passes, and hence implies a higher level.
The analysis revealed trivial effects of $T$ and $\sigma$.
Namely, the smaller $T$ or $\sigma$ was, i.e. the more competent OFs were, the greater the average number of passes was.
We note that $\sigma = 0$, which corresponds to no fluctuation in pass angles, was outside the value range in our analysis.
In our model, pass angle is measured relative to the vector connecting Passer to Receiver, as explained in '\modeldetail{}' in Methods section.
For instance, a pass whose angle is zero goes towards exactly where Receiver stands.
We expect the pass angle distribution to have non-zero variance, because pass angles would fluctuate due to not only mistakes but also strategic decisions.
That is, if Receiver stands close to DF, Passer is likely to give a pass in such a way that Receiver moves away from DF before obtaining the ball.
In contrast to $T$ and $\sigma$, $\tau$ and $\rmsub{L}{ict}$ should reflect the competence of DF.
The characteristic time scale of DF, $\tau$, affects the running speed of DF.
With a smaller value of $\tau$, DF runs faster, which would indicate more skilled DF.
Parameter $\rmsub{L}{ict}$ denotes the threshold distance for DF to intercept the ball.
Obviously, more experienced DF is likely to exhibit a larger value of $\rmsub{L}{ict}$.
Fig.~\ref{fig:sens.anal} illustrates that smaller $\tau$ or longer $\rmsub{L}{ict}$ reduced the average number of passes, which was consistent with our expectation.
In short, parameters $m$, $\gamma$, and $\rmsub{k}{r}$ turned out to have only limited effects on the number of passes as expected.
Moreover, parameters $T$, $\sigma$, $\tau$, and $\rmsub{L}{ict}$ exerted intuitively understandable influences.
These results imply the mechanistic plausibility of our model.

An interesting feature in Fig.~\ref{fig:sens.anal} is non-monotonicity, or optimality in other words.
This was salient for $\beta$ and $\rmsub{L}{e}$.
Inverse temperature $\beta$ controls Passer's accuracy in choosing the wider side when giving a pass.
The limiting case of $\beta = 0$ indicates complete randomness.
The larger the value of $\beta$ is, the more probable it is that Passer gives a pass to the player on the wider side. 
Parameter $\rmsub{L}{e}$ represents the typical distance of DF to Passer and to Mover.
The curves for $\beta$ and $\rmsub{L}{e}$ in Fig.~\ref{fig:sens.anal} suggest the existence of the optimal $\beta$ and $\rmsub{L}{e}$ values that maximise the average number of passes.
This implies that Passer should give a pass to the narrower side from time to time, and that Passer and Mover should maintain a certain distance from DF in our model.

\section*{Discussion}
With the aim of uncovering influence of football players' behaviour on team performance in 3v1 ball possession game, we constructed and analysed a dynamical model of the motion of the players and the ball.
We argued for the plausibility of our model by comparing the OF area and the average number of passes between experiments and simulations (Fig.~\ref{fig:random.params}).
Taking advantage of our modelling approach, the sensitivity analysis was performed to examine the influence of each model parameter on the number of passes, which allowed us to investigate the possible effects of interventions in behaviour.
Several parameters turned out to have intuitively understandable influences as expected, illustrating the mechanistic plausibility of our model.
We also found several crucial parameters that might reflect the competence of a team.

Through the sensitivity analysis, we obtained several implications about the way in which certain aspects of players' behaviour affect team performance in 3v1 ball possession game.
Firstly, faster and more accurate passes by OFs are expected to enhance ball possession, according to the results on $T$ and $\sigma$.
Secondly, the optimal, small randomness in choosing pass directions may improve ball possession.
This is seen from the result on $\beta$.
Whereas the ball should mostly go to the wider side (large $\beta$), giving passes to the narrower side from time to time is expected to improve the offensive team's performance.
Finally, the result on $\rmsub{L}{e}$ suggests that Passer and Mover can enhance ball possession by maintaining the optimal distance from DF.
We expect the optimal value of $\rmsub{L}{e}$ to depend on the size of the court.
In this study, the court was a square with $\qty{6}{\m}$ per side.
Therefore, if $\rmsub{L}{e}$ is set to around $\qty{6}{\m}$, which was longer than the optimal length according to our sensitivity analysis, Passer and Mover would always be running away from DF.
We conjecture that a too large value of $\rmsub{L}{e}$ degrades the formation of the offensive team by putting excessive weight on evasion of DF, undermining ball possession performance.
In addition to the sensitivity analysis, the process of constructing the model offered insights into the effect of DF's behaviour on the number of passes.
At first, we focused on modifying behaviour of OFs in the previous model~\cite{yokoyama2018Social}.
Then, it turned out that DF was hardly able to intercept the ball.
The average number of passes exceeded $1000$, when we turned off the maximum-pass termination condition.
This was how we learned that we needed to revise the behaviour not only of OFs but also of DF.
Since the previous model~\cite{yokoyama2018Social} incorporated only the `chase' motion mode of DF, we added the second motion mode of `wait', which allowed DF to intercept passes more frequently, considerably reducing the average number of passes.
In light of this experience, we presume it to be essential in ball possession dynamics that DF has the possibilities to both run and stay still.
In the previous agent-based model for 2v1 ball possession~\cite{chacoma2020Modeling}, while DF always moved towards the ball, the running speed was drawn from an exponential distribution at every time step.
Their choice for an exponential distribution could have been a key to their success, since its high probability density around $0$ would give rise to a motion similar to the `wait' mode.

Among the psychological literature on team effectiveness, it has been argued that more focus should be put on dynamic, complex, and emergent nature of team processes~\cite{kozlowski2018Unpacking,mathieu2019Embracing}.
To this end, Kozlowski and Chao~\cite{kozlowski2018Unpacking} described computational modelling as one of promising methodologies.
They discussed three potential difficulties for computational modeling, which we summarise as difficulties in
(1) constructing computational models based on theories expressed in natural languages, 
(2) balancing parsimony and complexity, and
(3) validating the models.
The first concern points to an inherent limitation of our approach.
That is, if one tries to incorporate existing insights into a model, they must be reformulated mathematically, through which the original nuance may well be lost.
This was the reason why we opted for physical analogy, such as elastic spring between players and viscosity.
Nevertheless, our study circumvented the second and third problems.
We concentrated on a specific case of 3v1 ball possession game, which made it easier to determine what needed to be modelled.
For instance, we incorporated two motion modes of DF, because it was the simplest choice to reproduce the empirical average number of passes.
In this way, we were able to keep our model from becoming too complex to interpret, addressing the second concern.
Finally, our model generates position data that are comparable with empirical tracking data.
As such, our model can be validated in terms of any quantity that is calculated from position data, including the OF area we showed in Fig.~\ref{fig:random.params}.
This significantly mitigates the third concern for verifiability.
We maintain that, to realise these advantages in studies using dynamical models, it is crucial to define a specific context, such as 3v1 ball possession game, and to simulate data that is as close to raw as possible, for instance position data from tracking systems in our case.

We owe the basic modelling framework, where a game is decomposed into pass scenes and transitions, to the previous work by Yokoyama et al.~\cite{yokoyama2018Social}.
Nevertheless, our model is distinct from the previous one because the motivation is different: while their interest lay in synchronisation patterns among OFs, our work concerns the number of passes.
Our major novelties include termination conditions, motion modes of DF, and the strategic choice of pass direction.
We also made technical modifications to enhance efficiency and stability.
For instance, the ODE system~[eq.~\eqref{eq:ode.system}] in our model can be integrated by numerical solvers with adaptive time steps, unlike the previous model, where the time step size of the Euler method, $\Delta t$, appears in the model equation.
In this article, we have discussed advantages of dynamical modelling for analysing influence of behaviour on team performance.
Accordingly, we hope to facilitate the use of dynamical models in future studies.
To this end, we aimed to present a model that is easy to modify, so that researchers can employ and adapt our model to suit their needs.
A salient possibility for revision lies in the social force term $\vec{G}$ in the equation of motion~\eqref{eq:PM.v.dot} of Passer and Mover.
One may even formulate social forces specific to each of Passer and Mover, which would come with the cost of the increase in the number of parameters.
Another possibility is to change the algorithm to determine the pass destination $\rmsub{\vec{d}}{B}^{n+1}$.
For instance, the pass angle $\theta_n$ may follow a distribution whose mean is not zero.
On the contrary, we would be discouraged to modify the confinement mechanism $\rmsup{\vec{F}}{return} + \rmsup{\vec{F}}{viscous}$ and DF's motion until it turns out necessary.
As we discuss in `\modeldetail{}' in Methods section, designing the confinement mechanism was challenging.
Moreover, details of the confinement mechanism should not influence the result as long as it works as expected.
Hence we do not see any reason for modification so far.
In this research, our focus is on the team performance of the offensive team. 
Therefore we regard DF as a part of the environment, which we wish to model in the as simple manner as possible, and we have not found the need for revision.

Although we were able to reproduce the average number of passes and OF area for the high- and low-level teams, there is still room for improvement in our model to enhance plausibility.
A closer look at the simulated dataset suggests that $\tau$ must to be small, i.e. DF must run faster, to mimic the empirical result of the low-level team.
This was contrary to our expectation that smaller $\tau$ would correspond to more competent DF.
It is probable that the OFs in our model are too competent and thus the only way to reduce the number of passes is to enhance the DF's capability.
A different kind of social forces which hinders the offensive team performance is apparently necessary to more accurately reproduce the dynamics of the low-level team.
There might be other discrepancies between the model and actual dynamics in ball possession game.
However, it is often unclear to what extent simulated dynamics should resemble reality, because the aim of modelling is extraction of the essential elements, rather than duplication of the reality.
In this work, our criterion for plausibility was based on the average number of passes and OF area.
The possible next step would be to revise the model to reproduce the distribution of the number of passes, not only the average.
Another limitation of our study is the choice of parameter values.
In the simulated dataset for the sensitivity analysis, we chose the value ranges of parameters manually.
Our decision could have affected the result.
For instance, a parameter that exhibited a monotonic effect in our analysis might have demonstrated a non-monotonic effect if one had opted for a wider value range.
However, there is no, at least practical, point in considering parameter values that are infeasible in reality.
If a certain amount of empirical data about focal teams are available, it should be possible to determine appropriate value ranges through data analysis.
We should highlight that these limitations should be borne in mind, especially when interpreting the result of the sensitivity analysis (Fig.~\ref{fig:sens.anal}).

We conclude by referring to three directions for future research.
Firstly, data assimilation between a model and the empirical data would be a natural next step, which we aim to work on.
In particular, after revising the model to overcome the limitations mentioned above, we aim to characterise the high- and low-level teams by values of model parameters and make comparison between them, which would allow us to empirically analyse determinants of high performance in ball possession.
For several parameters such as $T$, $m$, and $\rmsub{L}{e}$, empirically measured values can be used.
For parameters that cannot be directly measured, including $\beta$, $\rmsub{k}{e}$, and $\rmsub{k}{f}$, we need to conduct parameter estimation.
Once the differences between the high- and low-level teams are quantified, we should be able to design an experiment to verify the predictions.
A caveat is that our approach does not help us in modifying parameter values of actual teams: even though our study suggests an effect of pass accuracy on ball possession, how to improve pass accuracy of the low-level team is beyond the scope of our research.
Secondly, future studies can work on incorporating insights from sports science, especially performance analysis in football, into dynamical models.
This corresponds to the first of the three difficulties for computational modelling we summarised above.
Although translation of hypotheses expressed in natural languages into mathematical or computational formulations would be challenging, it would create novel opportunities to quantitatively verify knowledge from the existing literature.
Finally, our modelling approach is applicable to other settings regarding team effectiveness.
Although a model may take a completely different form, the basic procedure would be the same:
define state variables which are as close to raw data as possible,
describe their temporal evolution by mathematical models,
and discuss influences of model parameters on team effectiveness through sensitivity analysis.

\section*{Methods}
\subsection*{\experimentmethod{}}
\begin{figure}[tp]
  \centering
  \includegraphics[width=0.7\linewidth]{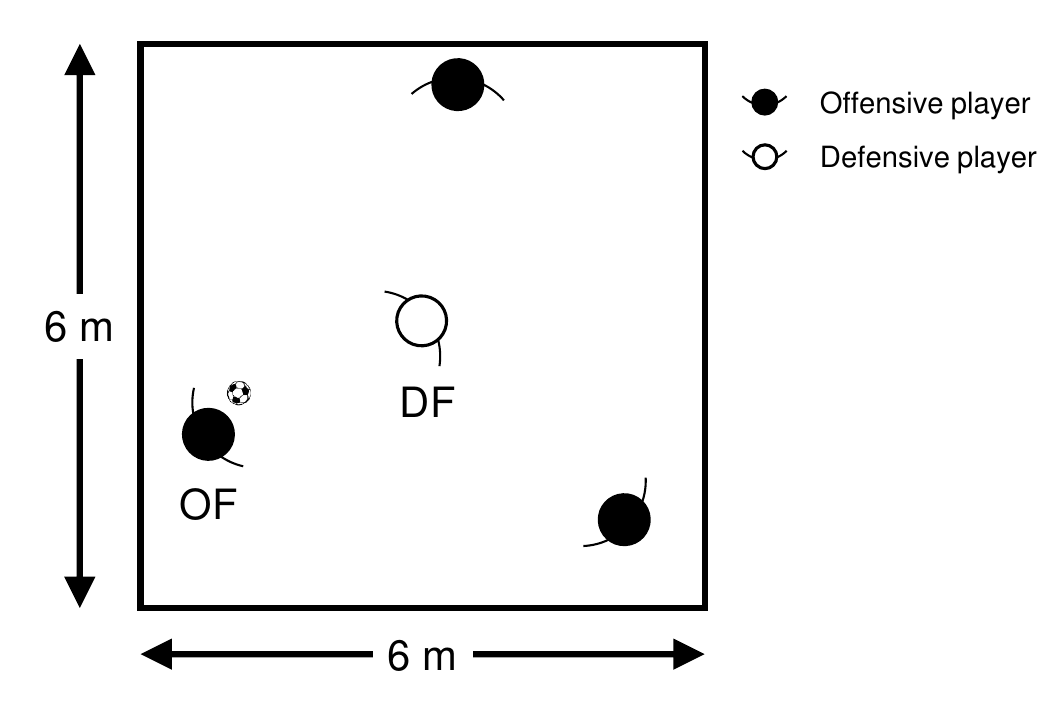}
  \caption{%
    Experimental setup of a ball possession training.
  }
  \label{fig:exp.setup}
\end{figure}

\paragraph{Participants}
Twelve male college football players were divided into two groups: the high-level group ($N = 5$; age: $22.2 \pm 0.3$ years, height: $176.6 \pm 4.3 \si{\cm}$, weight: $70.1 \pm 9.0 \si{\kg}$) and the low-level group ($N = 7$; age: $20.9 \pm 3.0$ years, height: $172.3 \pm 5.8 \si{\cm}$, weight: $69.3 \pm 5.3 \si{\kg}$). 
All participants had over 10 years of competitive experience. 
Participants trained according to their respective groups, engaging in six days of training per week, with each session lasting approximately two hours. 
All participants were healthy, without any muscular dysfunction or surgical/medical conditions. 
The purpose of the study and the risks associated with participation were thoroughly explained to the participants, and informed consent was obtained prior to the study. 
The study design was approved by an ethics review board (the Ethics Committee in National Institute of Fitness and Sports in Kanoya for human experimentation, \#11-101) and conducted in accordance with the Declaration of Helsinki.

\paragraph{Experimental design}
All measurements were conducted indoors on artificial turf (Grand Grass HD, Mizuno). 
For both groups, the task involved ball possession training within a square court with $\qty{6}{\m}$ on each side, with three offensive players (OFs) and one defensive player (DF), as depicted in Fig.~\ref{fig:exp.setup}.
OFs were instructed to complete 20 passes, with a maximum of two touches per player, while preventing DF from intercepting the ball. 
DF was tasked with intercepting the ball from OFs or kicking it out of the court before OFs could complete 20 consecutive passes. 
The trial ended when OFs completed 20 passes, DF intercepted the ball, or the ball went out of the bounds. 
At the end of each trial, one OF and DF were substituted, and the next trial commenced.

\paragraph{Measurement of players' location}
Position data on $(x, y)$-coordinate for each player were measured with local positioning system (ZXY Sports Tracking, Chyronhego, New York, USA) at a sampling frequency of $\qty{20}{\hertz}$. 
A belt with a sensor (approx. $\qty{20}{\g}$) was attached to each player. 
To prevent the belt from shifting during a trial, it was secured to their training wear with safety pins.

\paragraph{Calculation of the area formed by the three offensive players}
The area among the three OFs was calculated from the coordinates of the three OFs. 
The data were processed with Matlab (Mathworks R2021, New York, USA).
An exemplary video recording of the experiment and the visualised tracking data are available as \siexpvideo{}.

\subsection*{\modeldetail{}}
\begin{figure}[tp]
  \centering
  \includegraphics[width=0.8\linewidth]{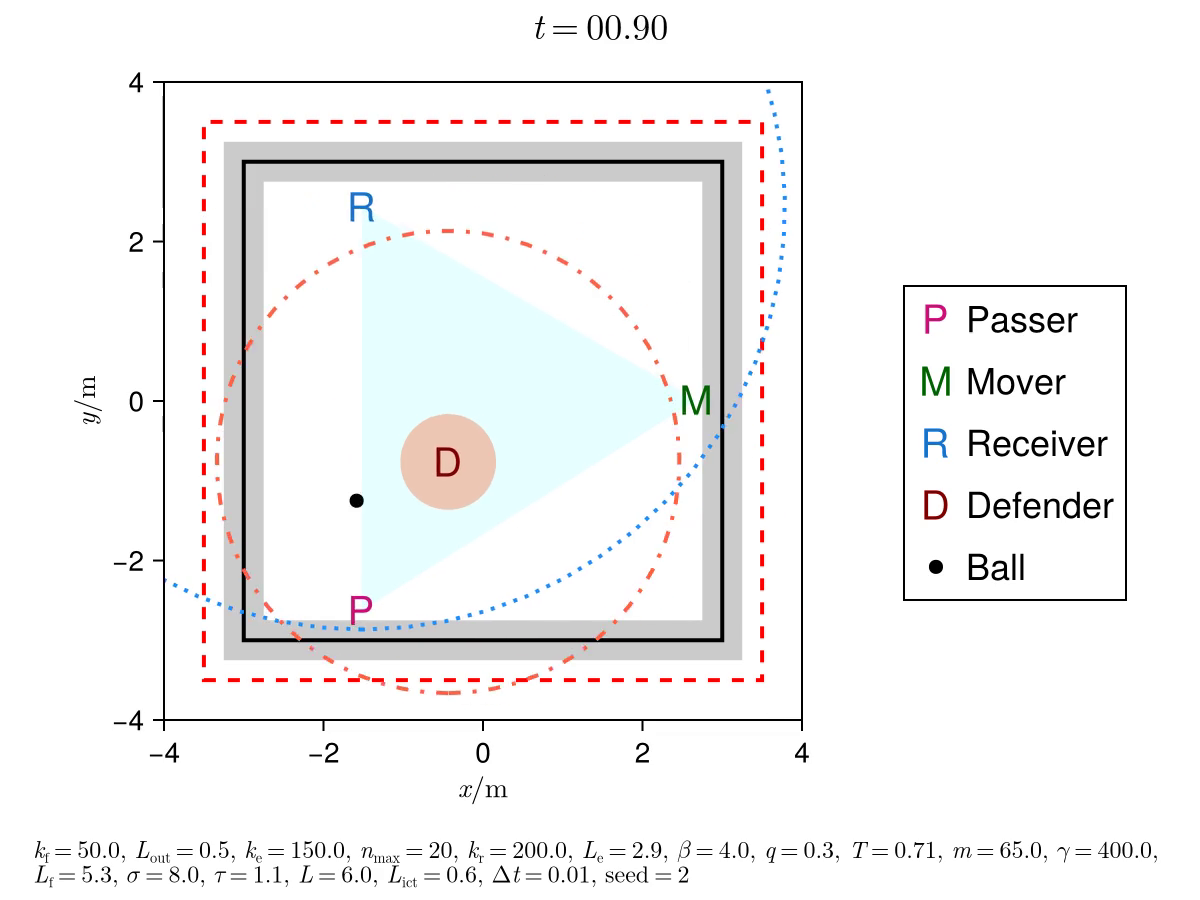}
  \caption{
    Snapshot of a simulated animation (\sisimvideowguide{}).
    It illustrates
      the court boundary (black solid line), 
      the threshold for ball-out event (red dashed line), 
      the OF area (blue triangle area), 
      the viscous region around the court boundary (gray area), 
      the intercept area around DF (red disk area), 
      the natural length of the evading force $\rmsub{L}{e}$ (orange dash-dotted line), 
      and the natural length of the following force $\rmsub{L}{f}$ (blue dotted line).
  }
  \label{fig:anim.snapshot}
\end{figure}

Step 2 (Fig.~\ref{fig:model.overview} b) in our model concerns the dynamics during a single pass scene.
We denote the positions of Receiver and the ball at the start of the $n$-th pass scene by $\rmsub{\vec{s}}{R}^n$ and $\rmsub{\vec{s}}{B}^n$, respectively.
We also introduce the destination of the $n$-th pass $\rmsub{\vec{d}}{B}^n$, i.e. the position of the ball when this pass scene ends as long as the game is not terminated.
During the $n$-th pass scene, the ball runs from $\rmsub{\vec{s}}{B}^n$ to $\rmsub{\vec{d}}{B}^n$.
Since Receiver must reach the ball at the end of this pass scene, Receiver moves from $\rmsub{\vec{s}}{R}^n$ to $\rmsub{\vec{d}}{B}^n$.
We note that the final positions of the ball and Receiver coincide.
For each pass scene, DF chooses one of the two motion mode: `chase' and `wait'.
When `wait', DF remains centred among OFs during the pass scene.
When `chase', on the other hand, DF attempts to intercept the pass by chasing the ball.
Passer and Mover can move more freely than other players, since they need not to reach the ball.
In the model, their motion is driven by social forces~\cite{helbing1995Social}, which describe a player's interactions with other players and environment.
A major constraint on their motion is that they must remain within, or at least in the vicinity of, the court during a game.
We designed two social forces to confine Passer and Mover within the court.
While the confinement mechanism must keep players from leaving the court, it should have little influence on the dynamics inside the court, since players are likely to care about the court boundary only when they are at the periphery.
This requirement made it challenging to design the confinement mechanism.
The dynamics of the five agents described so far are represented by the system of ODEs~\eqref{eq:ode.system}.

The motion of DF is governed by eq.~\eqref{eq:DF.x.dot}.
Binary parameter $\zeta_n \in \Bab{0, 1}$ represents the motion mode, with $\zeta_n = 0$ corresponding to `wait'.
The destination of DF at time $t$ is either $\rmsub{\vec{d}}{D,chase}(t)$ or $\rmsub{\vec{d}}{D,wait}(t)$.
When DF `chases' the ball, the destination is in the middle of the ball and Receiver:
\begin{gather}
  \rmsub{\vec{d}}{D, chase}(t) \coloneq \frac{\xR(t) + \xB(t)}{2}.
\end{gather}
When DF `waits' for an opportunity, the destination is the incentre of the triangle formed by OFs:
\begin{gather}
  \rmsub{\vec{d}}{D, wait}(t) \coloneq \frac{
    \norm{\xR(t) - \xM(t)} \xP(t) + \norm{\xP(t) - \xM(t)} \xR(t) + \norm{\xR(t) - \xP(t)} \xM(t)
  }{
    \norm{\xR(t) - \xM(t)} + \norm{\xP(t) - \xM(t)} + \norm{\xR(t) - \xP(t)}
  }.
\end{gather}

On the right-hand side of the dynamics for Passer and Mover [eq.~\eqref{eq:PM.v.dot}], social force terms appear.
First two terms correspond to the confinement mechanism.
One of the two is the returning force $\rmsup{\vec{F}}{return}$, which is an elastic attraction towards the court.
It acts independently along $x$ and $y$ axes, which are parallel to the court boundary (Fig.~\ref{fig:anim.snapshot}).
Along $x$ direction, the returning force is modelled by Hooke's law regarding the distance from the origin, which is at the centre of the court, with the natural length of $(L - \rmsub{L}{buff}) / 2$:
\begin{gather}
  \rmsup{F}{return}_x (x) \coloneq \begin{dcases}
    -\sign \pab{x} \rmsub{k}{r} \frac{\abs{x} - \pab{L - \rmsub{L}{buff}} / 2}{\rmsub{L}{buff} / 2}, 
      & \abs{x} \geq \frac{L - \rmsub{L}{buff}}{2} \\
    0, 
      & \abs{x} < \frac{L - \rmsub{L}{buff}}{2}.
  \end{dcases}
\end{gather}
The same applies along $y$ direction.
The natural length of $(L - \rmsub{L}{buff}) / 2$, not $L / 2$, reflects players' tendency to stand slightly within the court, not exactly on the boundary.
We note that the displacement is normalized by $\rmsub{L}{buff} / 2$.
We chose $\rmsub{L}{buff}$ to be $\qty{50}{\cm}$ throughout this study.
The other component of the confinement mechanism is the viscous force $\rmsup{\vec{F}}{viscous}$, which takes effect only in the neighbourhood of the court boundary (Fig.~\ref{fig:anim.snapshot}, gray area).
It acts on $x$ and $y$ directions independently as well, taking the form of nonlinear viscosity.
Along $x$ axis, it is described as
\begin{gather}
  \rmsup{F}{viscous}_x (x, v_x) \coloneq \begin{dcases}
    -\gamma \pab{v_x}^3, 
      & \abs{x} \in \bab{\frac{L - \rmsub{L}{buff}}{2}, \frac{L + \rmsub{L}{buff}}{2}} \\
    0, 
      & \text{otherwise},
  \end{dcases}
\end{gather}
where $v_x$ stands for $x$ component of the player's velocity.
The same holds along $y$ direction.
Buffer width $\rmsub{L}{buff}$ is the width of the viscous region, centred around the court boundary.
For a player's outward motion, the viscous force prevents the player from going out.
For the inward motion, it reduces the influence of the returning force on the dynamics inside the court.
The remaining term $\vec{G}$ can be any function of the state vector $\vec{u}_n$ and parameters $\vec{\theta}$ and $\vec{\lambda}_n$.
In our model, two social forces were included: the evading force $\rmsup{\vec{F}}{evade}$ and the following force $\rmsup{\vec{F}}{follow}$.
With the two terms, we incorporated dyadic interactions between OF and DF and among OFs.
The evading force $\rmsup{\vec{F}}{evade}$ models OF's tendency to stay away from DF.
It resembles a linear spring with the natural length of $\rmsub{L}{e}$ between DF and the focal player, who is either Passer or Mover (Fig.~\ref{fig:anim.snapshot}, orange dash-dotted line):
\begin{gather}
  \rmsup{\vec{F}}{evade} (\vec{x}, \xD) \coloneq -\rmsub{k}{e} \pab{\norm{\vec{x} - \xD} - \rmsub{L}{e}} \frac{\vec{x} - \xD}{\norm{\vec{x} - \xD}}.
\end{gather}
The following force $\rmsup{\vec{F}}{follow}$ aims to model OF's behaviour to prepare for the next pass scene.
As Receiver surely gives the next pass, we assumed other OFs would adjust their positions in relation to Receiver.
To keep our model simple, we approximated this behaviour by a linear spring with the natural length of $\rmsub{L}{f}$ (Fig.~\ref{fig:anim.snapshot}, blue dotted line):
\begin{gather}
  \rmsup{\vec{F}}{follow} (\vec{x}, \xR) \coloneq -\rmsub{k}{f} \pab{\norm{\vec{x} - \xR} - \rmsub{L}{f}} \frac{\vec{x} - \xR}{\norm{\vec{x} - \xR}}.
\end{gather}

Governing equations for Receiver, DF and the ball are first-order ODEs [eqs.~\eqref{eq:B.x.dot}, \eqref{eq:R.x.dot} and \eqref{eq:DF.x.dot}], which may be regarded as the overdamped limit of Newtonian equations of motion, so as to reduce the complexity of our model.
On the other hand, the governing equations for Passer and Mover [eqs.~\eqref{eq:PM.x.dot} and \eqref{eq:PM.v.dot}] take the form of Newtonian equation of motion.
This is because we are primarily interested in the offensive team performance, which we expect to be strongly affected by the off-the-ball movements of Passer and Mover.

At step 3 (Fig.~\ref{fig:model.overview} b), the termination conditions are evaluated.
If a condition is satisfied, the simulation ends.
We introduced three termination conditions: pass intercept, ball-out, and the maximum number of passes.
The $n$-th pass is considered to be intercepted, if the ball has been closer to DF than the threshold $\rmsub{L}{ict}$ during $t_n \in [0, T]$.
Similarly, we say the ball went out of the court during the $n$-th pass if the ball went further than the threshold distance $\rmsub{L}{out}$ from the court during $t_n \in [0, T]$.
In these cases, the number of successful passes of this game would be $n - 1$.
Finally, the simulation is terminated when the number of passes reaches $20$ in accordance with the experiment's setting.

Step 4 concerns the transition from the $n$-th to $(n + 1)$-th pass scene.
Pass-specific parameters and initial conditions must be determined based on the state at the end of the previous pass scene.
First, roles among OFs switch.
Obviously, Receiver of the $n$-th pass becomes Passer of the $(n + 1)$-th pass.
Between the remaining two OFs, we postulate that the one with a wider pass course is more likely to become the next Receiver.
This feature is implemented by defining the probability that the current Passer becomes the next Receiver as
\begin{gather}
  p_{\mathrm{P}\to\mathrm{R}} \coloneq \frac{
    \exp \pab{\beta \angle \mathrm{PRD}}
  }{\exp \pab{\beta \angle \mathrm{MRD}} + \exp \pab{\beta \angle \mathrm{PRD}}},
\end{gather}
where inverse temperature $\beta \geq 0$ controls the accuracy of choosing the wider side.
The limit of high temperature, $\beta = 0$, indicates a complete randomness, and $\beta \to \infty$ represents a deterministic choice for the one with the wider angle.
Once the next roles among OFs are determined, the initial conditions for their positions are fixed.
For example, Passer of the $(n+1)$-th pass scene must start from the position where the previous Receiver, who is the same individual, stood at the last moment in the $n$-th pass scene.
To take into account the drastic reorientation of players, initial velocities are set to zero for all players, i.e. $\vec{v}_i(t_{n+1} = 0) = \vec{0}$.
In the model, the destination of the $(n+1)$-th pass $\rmsub{\vec{d}}{B}^{n+1}$ is determined before the $(n+1)$-th pass scene starts.
Although the ball runs generally towards Receiver, its direction can deviate either intentionally or by mistake.
Therefore, similarly to the pervious work~\cite{yokoyama2018Social}, $\rmsub{\vec{d}}{B}^{n+1}$ is calculated by rotating the vector $\xR^* - \xP^*$, which directs from Passer to Receiver of the $n$-th pass at the moment of the transition, around $\xP^*$ by a random angle of $\theta_n$ drawn from a normal distribution $\mathcal{N}(0, \sigma^2)$.
The DF's motion mode depends stochastically on the previous one.
With the probability of $q$, which is a model parameter, DF employs the same motion mode as in the previous pass scene.
An extreme case of $q = 0$ means that the DF's motion mode alternates between `wait' and `chase', while $q = 1$ indicates that DF keep choosing the same mode throughout a game.

During each pass scene, dynamics are governed by the ODEs, and therefore dynamics are deterministic.
The transition algorithm is a stochastic mapping from the $n$-th to $(n + 1)$-th pass scene.
Hence, it may be possible to regard our model as a Markov process by considering each pass scene to be a single `state'.
This class of models differs from other well-known classes of stochastic models such as stochastic differential equations or random ODEs.
It may be beneficial to elaborate on mathematical properties of this class of models, especially in relation to development of data assimilation methodologies.

\subsection*{\nummethod{}}
Table~\ref{tab:param.list} lists the model parameters.
Simulations and analyses in this study were performed with Julia.
We utilised DifferentialEquations.jl package~\cite{rackauckas2017differentialequations} for the numerical integration of the ODE system [eq.~\eqref{eq:ode.system}].

\subsection*{\datagen{}}
To prepare Fig.~\ref{fig:random.params} and perform the sensitivity analysis, we generated a dataset composing of randomly sampled values of model parameters, the average number of passes, and the average OF area.
Among the parameters listed in table~\ref{tab:param.list}, we fixed the values of $L$, $\rmsub{L}{buff}$, and $\rmsub{L}{out}$.
For each of the other 13 model parameters, a value was drawn from a uniform distribution whose range is indicated in table~\ref{tab:param.list}.
We determined the value ranges in consideration of the previous study~\cite{yokoyama2018Social} and plausibility.
For each set of parameter values, 500 games were simulated.
We recorded the number of passes and the temporal average of OF area for every game, and then took ensemble average over 500 trials to obtain the average number of passes and the temporal and ensemble average of OF area, which we simply denote as the average OF area, for each set of parameter values.
This procedure was repeated to obtain data for $15,000$ sets of parameter values.
The simulated dataset is visually summarized in \sihistorg{}.

A small portion of parameter values induced unrealistic results in terms of the average number of passes and the average OF area.
Since our major concern was on influences of model parameters on the outcomes, not on conditions for the model to work properly in the first place, we excluded unreasonable data points according to the following criteria.
Firstly, the average number of passes must not be less than $1$.
In addition, the average OF area must be less than $\qty{18}{m^2}$, which is the area of the largest triangle within the court.
Moreover, the number of games terminated due to the ball-out event must not be more than $400$ out of $500$ trials.
As a result, $895$ samples were excluded, leaving $14,105$ samples for further analyses.
The dataset after the preprocessing is visually summarised in \sihistpro{}.

\subsection*{\sensanal{}}
Using the preprocessed dataset, we performed multivariate quadratic regression to examine effects of parameters that were varied.
The regression model was
\begin{gather} \label{eq:quadratic.model}
  y_i = a_i + \sum_{k=1}^{13} \pab{b_k x_i^{(k)} + c_k \bab{x_i^{(k)}}^2 + \sum_{l>k} d_{kl} \bab{x_i^{(k)} \cdot x_i^{(l)}}} + \epsilon_i,
\end{gather}
where $i$ is the index of sets of parameters, $y_i$ is the average number of passes, $x_i^{(k)}$ is the standardised value of the $k$-th parameter whose mean is $0$ and variance is $1$, and $\epsilon_i$ is the residual following a normal distribution.
On the right-hand side, $a_i$ denotes the intercept, $b_k$ and $c_k$ represent the linear and square effects of the $k$-th parameter, respectively, and $d_{kl}$ indicates the interaction effect between the $k$-th and $l$-th parameters.
In Fig.~\ref{fig:sens.anal}, we plotted $\hat{b}_k x^{(k)} + \hat{c}_k \bab{x^{(k)}}^2$, where $\hat{b}_k$ and $\hat{c}_k$ were the estimated coefficient values and $x^{(k)}$ was the standardised value of the $k$-th parameter.
We replaced $\hat{b}_k$ or $\hat{c}_k$ by zero when it was not statistically significant with the significance level of $0.1\%$.
Parameter $\gamma$ was omitted in Fig.~\ref{fig:sens.anal} because neither coefficients were significant.
In \siregtable{}, we report the regression table showing the estimated coefficients and their standard errors.
The table also contains the results for the linear model, which is the case of $c_k \equiv 0$ and $d_{kl} \equiv 0$, and the square model without interaction terms, which is the case of $d_{kl} \equiv 0$.
The full quadratic model~[eq.~\eqref{eq:quadratic.model}] was superior to the linear and the square models in terms of adjusted $R^2$, AIC, BIC, and $F$-test, as shown in \siregcomp{}.

\section*{Data Availability}
The Julia codes for the simulation and the analyses, the simulated dataset utilised in the sensitivity analysis, and the empirical data on the number of passes and the average OF area are available in the GitHub repository, \url{https://github.com/ishiihidemasa/25-dynamical-modelling-3v1-ball-possession.git}.

\bibliographystyle{unsrt}
\bibliography{24-baseline-model-ball-possession}

\section*{Acknowledgements}
H.I. benefited from discussions at the CRC International Summer School 2024 of SFB1294 Data Assimilation, and from valuable comments from Toshiaki Asakura.
This study was partly supported by the World-leading Innovative Graduate Study Program in Proactive Environmental Studies (WINGS-PES), the University of Tokyo, and JSPS KAKENHI Grant Number JP24KJ0635 to H.I.

\section*{Author contributions statement}
H.I. conceived the study, developed the model, conducted the numerical simulations and analyses, and wrote the manuscript.
Y.T. designed and conducted the experiment, processed the obtained tracking data, and wrote the manuscript.
All authors contributed to the model development and its analyses throughout the research process and reviewed the manuscript.

\section*{Additional information}
\paragraph*{Competing interests} The authors declare no competing interests.
\end{document}


\maketitle
\captionsetup{width=.9\linewidth}

\section*{Simulated dataset before and after preprocessing}
\begin{figure}[tb]
  \centering
  \includegraphics[width=0.96\linewidth]{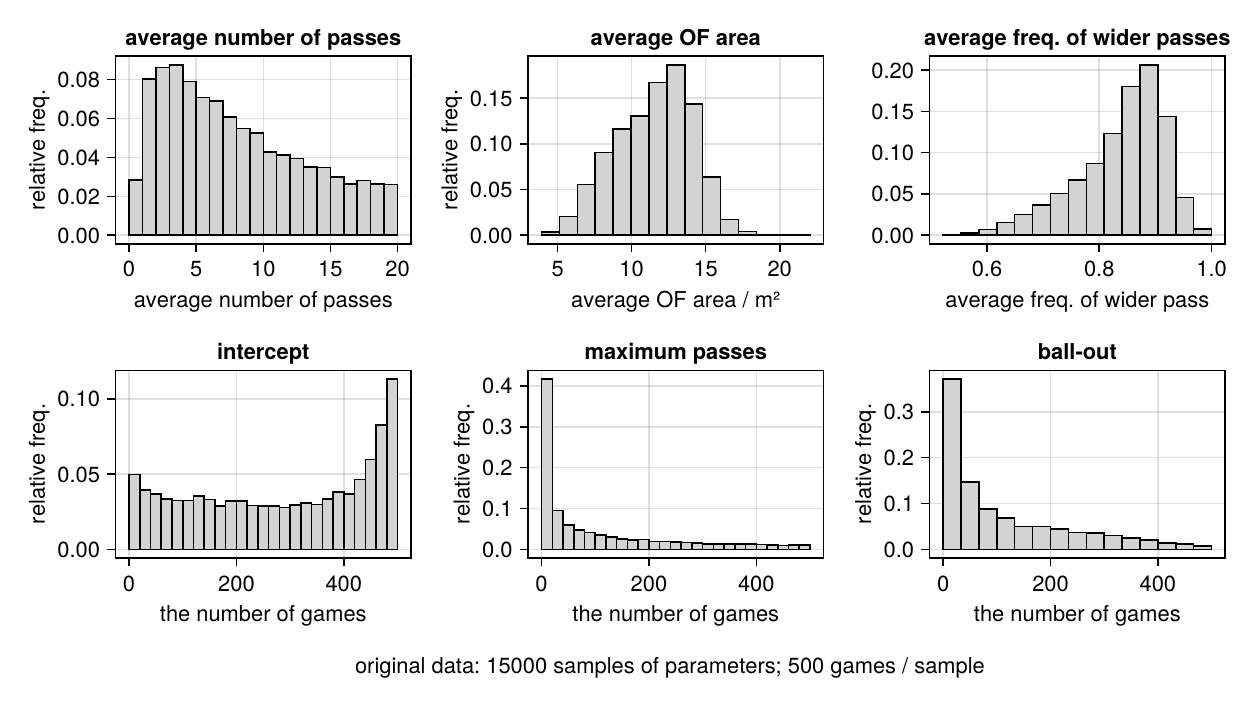}
  \caption{%
    Simulated dataset before preprocessing.
  }
  \label{fig:sim.hist.org}
\end{figure}

\begin{figure}[tb]
  \centering
  \includegraphics[width=0.96\linewidth]{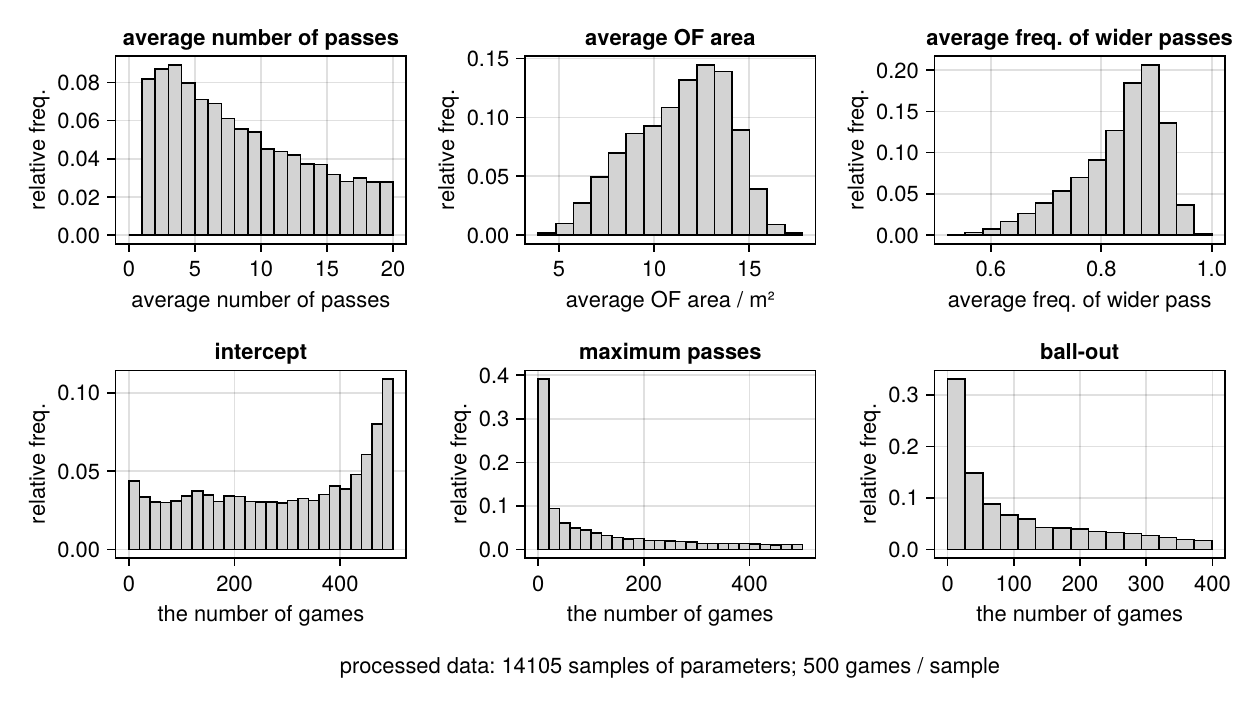}
  \caption{%
    Simulated dataset after preprocessing.
  }
  \label{fig:sim.hist.pro}
\end{figure}

In figures~\ref{fig:sim.hist.org} and \ref{fig:sim.hist.pro}, the simulated dataset before and after preprocessing are summarised.
In each figure, the upper three panels show the histograms of the average number of passes, the ensemble and temporal average of OF area, and the average relative frequency of passes that went to the wider side.
The lower three panels indicated the relative frequency among $500$ simulated games of the events that terminated the simulated game.
The figure indicates the limited influence of the preprocessing on the distributions.

\section*{Supplementary Videos}
\begin{description}
  \item[S1] Simulated game terminated by pass intercept shown with visual guide.
  \item[S2] Simulated game terminated by the maximum number of passes shown without guide.
  \item[S3] Simulated game terminated by ball-out event shown without guide.
  \item[S4] Simulated game terminated by pass intercept shown without guide.
  \item[S5] Video recording of the experiment with the high-level team (left) and its position data (right).
\end{description}

\FloatBarrier
\section*{Details on the regression results}
We report in Supplementary Table~\ref{tab:regression.table} the regression table indicating the estimated values of coefficients and their standard errors.
Interaction effects in the quadratic model are visualised in Supplementary Fig.~\ref{fig:interaction.effect}.
Further details on each model, including $t$-statistics and $p$ values, can be found in the GitHub repository \url{https://github.com/ishiihidemasa/25-dynamical-modelling-3v1-ball-possession.git} together with the codes for analyses.
The dependent variable was the average number of passes over $500$ simulated games.
The independent variables, which were parameter values, were centred around the sample mean within the dataset.
For the quadratic model, the independent variables were further standardized to have the standard deviation of one.
Therefore, there is no point in comparing the absolute coefficient values of the quadratic model to the ones of the linear and the square model.

In the main article, we discussed only the quadratic model.
This was because it performed the best in terms of the adjusted $R^2$, AIC, BIC, and $F$-test, as indicated in Supplementary Table~\ref{tab:reg.comp}.

\begin{figure}[tbh]
  \centering
  \includegraphics[width=0.7\linewidth]{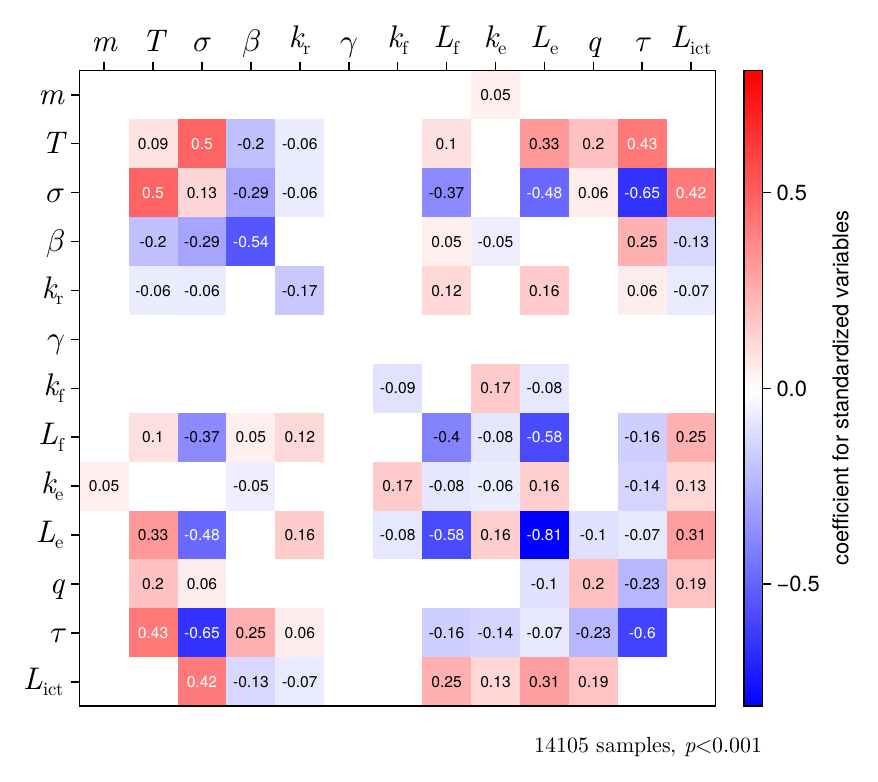}
  \caption{
    Interaction effects among parameters.
    Significant interaction effects among the parameters with the significance level of $0.1\%$.
    Estimated coefficients are both labelled within cells and illustrated by color.
  }
  \label{fig:interaction.effect}
\end{figure}

\clearpage
\begin{longtable}{rrrr}
  \caption{
    Regression table for the linear, square (without interaction terms), and quadratic models.
    *, **, and *** represent significance levels of 5\%, 1\%, and 0.1\%, respectively.
  }
  \label{tab:regression.table}
  \\
  \toprule
                         &  \multicolumn{3}{c}{effect on the number of passes} \\ 
  \cmidrule(lr){2-4} 
                         &     linear &    square &  quadratic \\ 
  \midrule
  \endhead
  (Intercept)            &   8.379*** &  10.334*** &  10.565*** \\ 
                         &    (0.022) &    (0.082) &    (0.063) \\ 
  $m$                    &    0.011** &     0.008* &   0.053*** \\ 
                         &    (0.004) &    (0.003) &    (0.015) \\ 
  $T$                    & -10.575*** & -10.678*** &  -2.181*** \\ 
                         &    (0.107) &    (0.097) &    (0.015) \\ 
  $\sigma$               &  -1.303*** &  -1.304*** &  -2.295*** \\ 
                         &    (0.012) &    (0.011) &    (0.015) \\ 
  $\beta$                &   0.129*** &   0.128*** &   0.904*** \\ 
                         &    (0.003) &    (0.003) &    (0.015) \\ 
  $\kr$                  &   0.001*** &   0.001*** &   0.227*** \\ 
                         &    (0.000) &    (0.000) &    (0.015) \\ 
  $\gamma$               &      0.000 &     0.000* &    0.046** \\ 
                         &    (0.000) &    (0.000) &    (0.015) \\ 
  $\kf$                  &      0.000 &      0.000 &     -0.022 \\ 
                         &    (0.000) &    (0.000) &    (0.015) \\ 
  $\Lf$                  &   0.120*** &   0.129*** &   0.164*** \\ 
                         &    (0.015) &    (0.014) &    (0.015) \\ 
  $\ke$                  &  -0.005*** &  -0.006*** &  -0.363*** \\ 
                         &    (0.000) &    (0.000) &    (0.015) \\ 
  $\Le$                  &   0.431*** &   0.446*** &   0.612*** \\ 
                         &    (0.015) &    (0.014) &    (0.015) \\ 
  $q$                    &   1.484*** &   1.624*** &   0.309*** \\* 
                         &    (0.126) &    (0.114) &    (0.015) \\ 
  $\tau$                 &   8.332*** &   8.242*** &   2.361*** \\* 
                         &    (0.076) &    (0.069) &    (0.015) \\ 
  $\Lict$                & -19.320*** & -19.415*** &  -2.243*** \\* 
                         &    (0.188) &    (0.171) &    (0.015) \\ 
  $m \times m$           &            &      0.000 &     -0.006 \\* 
                         &            &    (0.001) &    (0.017) \\ 
  $T \times T$           &            &   3.074*** &   0.094*** \\ 
                         &            &    (0.539) &    (0.017) \\ 
  $\sigma \times \sigma$ &            &   0.040*** &   0.131*** \\ 
                         &            &    (0.007) &    (0.017) \\ 
  $\beta \times \beta$   &            &  -0.011*** &  -0.539*** \\ 
                         &            &    (0.000) &    (0.017) \\ 
  $\kr \times \kr$       &            &  -0.000*** &  -0.174*** \\ 
                         &            &    (0.000) &    (0.017) \\ 
  $\gamma \times \gamma$ &            &     -0.000 &      0.000 \\ 
                         &            &    (0.000) &    (0.017) \\ 
  $\kf \times \kf$       &            &    -0.000* &  -0.094*** \\ 
                         &            &    (0.000) &    (0.017) \\ 
  $\Lf \times \Lf$       &            &  -0.184*** &  -0.396*** \\ 
                         &            &    (0.011) &    (0.017) \\ 
  $\ke \times \ke$       &            &    -0.000* &  -0.060*** \\ 
                         &            &    (0.000) &    (0.017) \\ 
  $\Le \times \Le$       &            &  -0.376*** &  -0.813*** \\ 
                         &            &    (0.011) &    (0.017) \\ 
  $q \times q$           &            &   7.649*** &   0.202*** \\* 
                         &            &    (0.737) &    (0.017) \\ 
  $\tau \times \tau$     &            &  -6.464*** &  -0.602*** \\* 
                         &            &    (0.267) &    (0.017) \\ 
  $\Lict \times \Lict$   &            &      1.886 &     -0.012 \\* 
                         &            &    (1.655) &    (0.017) \\ 
  $T \times m$           &            &            &    -0.038* \\* 
                         &            &            &    (0.015) \\ 
  $\sigma \times m$      &            &            &     -0.017 \\* 
                         &            &            &    (0.015) \\ 
  $\beta \times m$       &            &            &      0.006 \\* 
                         &            &            &    (0.015) \\ 
  $\kr \times m$         &            &            &     -0.004 \\* 
                         &            &            &    (0.015) \\ 
  $\gamma \times m$      &            &            &    -0.031* \\ 
                         &            &            &    (0.015) \\ 
  $\kf \times m$         &            &            &     -0.012 \\ 
                         &            &            &    (0.015) \\ 
  $\Lf \times m$         &            &            &      0.024 \\ 
                         &            &            &    (0.015) \\ 
  $\ke \times m$         &            &            &   0.052*** \\ 
                         &            &            &    (0.015) \\ 
  $\Le \times m$         &            &            &   -0.045** \\ 
                         &            &            &    (0.015) \\ 
  $q \times m$           &            &            &      0.008 \\ 
                         &            &            &    (0.015) \\ 
  $\tau \times m$        &            &            &      0.014 \\* 
                         &            &            &    (0.015) \\ 
  $\Lict \times m$       &            &            &     -0.017 \\* 
                         &            &            &    (0.015) \\ 
  $\sigma \times T$      &            &            &   0.495*** \\* 
                         &            &            &    (0.015) \\ 
  $\beta \times T$       &            &            &  -0.202*** \\* 
                         &            &            &    (0.015) \\ 
  $\kr \times T$         &            &            &  -0.061*** \\* 
                         &            &            &    (0.015) \\ 
  $\gamma \times T$      &            &            &    0.046** \\* 
                         &            &            &    (0.015) \\ 
  $\kf \times T$         &            &            &    -0.030* \\* 
                         &            &            &    (0.015) \\ 
  $\Lf \times T$         &            &            &   0.100*** \\* 
                         &            &            &    (0.015) \\ 
  $\ke \times T$         &            &            &     -0.013 \\* 
                         &            &            &    (0.015) \\ 
  $\Le \times T$         &            &            &   0.325*** \\* 
                         &            &            &    (0.015) \\ 
  $q \times T$           &            &            &   0.198*** \\* 
                         &            &            &    (0.015) \\ 
  $\tau \times T$        &            &            &   0.426*** \\* 
                         &            &            &    (0.015) \\ 
  $\Lict \times T$       &            &            &      0.018 \\* 
                         &            &            &    (0.015) \\ 
  $\beta \times \sigma$  &            &            &  -0.287*** \\* 
                         &            &            &    (0.015) \\ 
  $\kr \times \sigma$    &            &            &  -0.065*** \\* 
                         &            &            &    (0.015) \\ 
  $\gamma \times \sigma$ &            &            &     -0.008 \\* 
                         &            &            &    (0.015) \\ 
  $\kf \times \sigma$    &            &            &     -0.026 \\* 
                         &            &            &    (0.015) \\ 
  $\Lf \times \sigma$    &            &            &  -0.369*** \\* 
                         &            &            &    (0.015) \\ 
  $\ke \times \sigma$    &            &            &      0.019 \\* 
                         &            &            &    (0.015) \\ 
  $\Le \times \sigma$    &            &            &  -0.479*** \\* 
                         &            &            &    (0.015) \\ 
  $q \times \sigma$      &            &            &   0.059*** \\* 
                         &            &            &    (0.015) \\ 
  $\tau \times \sigma$   &            &            &  -0.651*** \\* 
                         &            &            &    (0.015) \\ 
  $\Lict \times \sigma$  &            &            &   0.424*** \\* 
                         &            &            &    (0.015) \\ 
  $\kr \times \beta$     &            &            &      0.027 \\* 
                         &            &            &    (0.015) \\ 
  $\gamma \times \beta$  &            &            &     -0.001 \\* 
                         &            &            &    (0.015) \\ 
  $\kf \times \beta$     &            &            &     0.035* \\* 
                         &            &            &    (0.015) \\ 
  $\Lf \times \beta$     &            &            &   0.052*** \\* 
                         &            &            &    (0.015) \\ 
  $\ke \times \beta$     &            &            &  -0.054*** \\* 
                         &            &            &    (0.015) \\ 
  $\Le \times \beta$     &            &            &     -0.010 \\* 
                         &            &            &    (0.015) \\ 
  $q \times \beta$       &            &            &      0.009 \\* 
                         &            &            &    (0.015) \\ 
  $\tau \times \beta$    &            &            &   0.253*** \\* 
                         &            &            &    (0.015) \\ 
  $\Lict \times \beta$   &            &            &  -0.127*** \\* 
                         &            &            &    (0.015) \\ 
  $\gamma \times \kr$    &            &            &      0.011 \\* 
                         &            &            &    (0.015) \\ 
  $\kf \times \kr$       &            &            &      0.021 \\* 
                         &            &            &    (0.015) \\ 
  $\Lf \times \kr$       &            &            &   0.123*** \\* 
                         &            &            &    (0.015) \\ 
  $\ke \times \kr$       &            &            &    0.041** \\* 
                         &            &            &    (0.015) \\ 
  $\Le \times \kr$       &            &            &   0.165*** \\* 
                         &            &            &    (0.015) \\ 
  $q \times \kr$         &            &            &      0.027 \\* 
                         &            &            &    (0.015) \\ 
  $\tau \times \kr$      &            &            &   0.057*** \\* 
                         &            &            &    (0.015) \\ 
  $\Lict \times \kr$     &            &            &  -0.067*** \\* 
                         &            &            &    (0.015) \\ 
  $\kf \times \gamma$    &            &            &     -0.016 \\* 
                         &            &            &    (0.015) \\ 
  $\Lf \times \gamma$    &            &            &   -0.046** \\* 
                         &            &            &    (0.015) \\ 
  $\ke \times \gamma$    &            &            &      0.025 \\* 
                         &            &            &    (0.015) \\ 
  $\Le \times \gamma$    &            &            &     -0.022 \\* 
                         &            &            &    (0.015) \\ 
  $q \times \gamma$      &            &            &     -0.014 \\* 
                         &            &            &    (0.015) \\ 
  $\tau \times \gamma$   &            &            &     -0.003 \\* 
                         &            &            &    (0.015) \\ 
  $\Lict \times \gamma$  &            &            &      0.006 \\* 
                         &            &            &    (0.015) \\ 
  $\Lf \times \kf$       &            &            &     -0.013 \\* 
                         &            &            &    (0.015) \\ 
  $\ke \times \kf$       &            &            &   0.168*** \\* 
                         &            &            &    (0.015) \\ 
  $\Le \times \kf$       &            &            &  -0.077*** \\* 
                         &            &            &    (0.015) \\ 
  $q \times \kf$         &            &            &     -0.012 \\* 
                         &            &            &    (0.015) \\ 
  $\tau \times \kf$      &            &            &      0.005 \\* 
                         &            &            &    (0.015) \\ 
  $\Lict \times \kf$     &            &            &     -0.003 \\* 
                         &            &            &    (0.015) \\ 
  $\ke \times \Lf$       &            &            &  -0.081*** \\* 
                         &            &            &    (0.015) \\ 
  $\Le \times \Lf$       &            &            &  -0.577*** \\* 
                         &            &            &    (0.015) \\ 
  $q \times \Lf$         &            &            &      0.017 \\* 
                         &            &            &    (0.015) \\ 
  $\tau \times \Lf$      &            &            &  -0.157*** \\* 
                         &            &            &    (0.015) \\ 
  $\Lict \times \Lf$     &            &            &   0.250*** \\* 
                         &            &            &    (0.015) \\ 
  $\Le \times \ke$       &            &            &   0.156*** \\* 
                         &            &            &    (0.015) \\ 
  $q \times \ke$         &            &            &     -0.026 \\* 
                         &            &            &    (0.015) \\ 
  $\tau \times \ke$      &            &            &  -0.136*** \\* 
                         &            &            &    (0.015) \\ 
  $\Lict \times \ke$     &            &            &   0.128*** \\* 
                         &            &            &    (0.015) \\ 
  $q \times \Le$         &            &            &  -0.098*** \\* 
                         &            &            &    (0.015) \\ 
  $\tau \times \Le$      &            &            &  -0.075*** \\* 
                         &            &            &    (0.015) \\ 
  $\Lict \times \Le$     &            &            &   0.310*** \\* 
                         &            &            &    (0.015) \\ 
  $\tau \times q$        &            &            &  -0.231*** \\* 
                         &            &            &    (0.015) \\ 
  $\Lict \times q$       &            &            &   0.190*** \\* 
                         &            &            &    (0.015) \\ 
  $\Lict \times \tau$    &            &            &     0.034* \\* 
                         &            &            &    (0.015) \\ 
  \midrule
  $N$                    &     14,105 &     14,105 &     14,105 \\ 
  $R^2$                  &      0.758 &      0.800 &      0.882 \\ 
  \bottomrule
\end{longtable}

\begin{table}[tbh]
  \centering
  \caption{The quadratic model performed better than the other two models.
    $F$-statistic and $\mathrm{Prob}([F, \infty))$ summarize the results of $F$-tests
    comparing the square model to the linear one and the quadratic model to the square one.
  }
  \label{tab:reg.comp}
  \begin{tabular}{lrrrrr}
    \toprule
                    & adjusted $R^2$ & AIC & BIC & $F$-statistic & $\mathrm{Prob}([F, \infty))$ \\
    \midrule
    linear model    & 0.7577 & 66642.3234 & 66755.6377 &          &               \\
    square model   & 0.7997 & 63969.8057 & 64181.3257 & 228.3262 & $< 10^{-99}$  \\
    quadratic model & 0.8813 & 56669.9435 & 57470.6977 & 125.0219 & $< 10^{-99}$  \\
    \bottomrule
  \end{tabular}
\end{table}